\documentclass[11pt]{article}
\title{\bf An alternative to quintessence}

\author {Alexander Kamenshchik$^1$, Ugo Moschella$^2$ and  Vincent
Pasquier$^3$\\[10pt]
$^1$L. D. Landau Institute for Theoretical Physics,\\
Russian Academy of Sciences, \\
2 Kosygina street, 117334, Moscow,
Russia\\[3pt]
Landau Network -- Centro Volta, Villa Olmo, \\
via Cantoni 1, 22100 Como, Italy\\
 $^2$Dipartimento di Scienze Matematiche Fisiche e Chimiche, \\
 Universit\`a dell Insubria, \\
 Via Valleggio 11
 , 22100 Como \\ and INFN sez. di Milano, Italy\\[3pt]
 $^3$Service de Physique Th\'eorique, C.E. Saclay,\\
91191 Gif-sur-Yvette, France}
\begin{document}
\maketitle
\begin{abstract}
We consider a FRW cosmological model with an exotic fluid known
as Chaplygin gas. We show that the resulting evolution of the
universe is not in disagreement with the current observation of
cosmic acceleration. The model predict an increasing value for
the effective cosmological constant.
\end{abstract}

The discovery that the  expansion of the universe is accelerating
\cite{accel} has promoted the search for new types of matter that
can behave like a cosmological constant
\cite{Caldwell:1998ii,Sahni} by combining positive energy density
and negative pressure. This type of matter is often called
''quintessence''.

Since in a variety of inflationary models scalar fields have been
used in describing the transition from the quasi-exponential
expansion of the early universe to a power law expansion, it is
natural to try to understand the present acceleration of the
universe, which has an exponential behaviour too, by constructing

models where the matter responsible for such behaviour is also
represented by a scalar field \cite{Star}. However, now we deal
with the opposite task, i.e. we would like to describe the
transition from a universe filled with dust-like matter to an
exponentially expanding universe, and scalar fields are not the
only possibility but there are (of course) alternatives. In
particular, one can try to do it by using some perfect fluid but
obeying ``exotic'' equations of state.

In this short letter we consider an example  of this type: the
so-called {\em Chaplygin gas}. Under this name we mean a perfect
fluid having the following equation of state:
\begin{equation}
p = - \frac{A}{\rho},
\label{Chapl}
\end{equation}
where $p$ and  $\rho$ are respectively pressure and energy
density in a comoving reference frame, with $\rho>0$; $A$ is a
positive constant. This equation of state has raised recently a
certain interest \cite{Chapl} because of its many interesting
and, in some sense, intriguingly unique features. Indeed Eq.
(\ref{Chapl}) has an amusing connection with string theory and it
can be obtained from the Nambu-Goto action for $d$-branes moving
in a $(d+2)$-dimensional spacetime in the light-cone
parametrization \cite{Chaply}. Also, the Chaplygin gas is the
only fluid which, up to now, admits a supersymmetric
generalization \cite{superHoppe,jackiw}. We ourselves came across
this fluid \cite{we} when studying the stabilization of branes
\cite{RS} in black hole bulks \cite{BHTZ}. We found that to obtain
stabilization it is necessary  to add matter on the branes which
again obeys the equation of state (\ref{Chapl}).

For these reasons we have undertaken the simple exercise of
stydying  a FRW cosmology of a universe filled with a Chaplygin
gas. The metric of a homogeneous and isotropic universe is
usually written as follows
\begin{equation}
ds^2 = dt^2 - a^2(t)dl^2, \label{Fried}
\end{equation}
where $dl^2$ is the metric of a 3-manifold of constant curvature
($K= 0,\pm 1$), and the expansion factor $a(t)$ evolves according
to the Friedmann equation
\begin{equation}
\frac{\dot{a}^2}{a^2} = \rho - \frac{K}{a^2}. \label{Fried1}
\end{equation}

Energy conservation
\begin{equation}
d(\rho \, a^3) = - p\ d (a^3) \label{Fried2}
\end{equation}
 together with the equation of state (\ref{Chapl}) give the following relation:
\begin{equation}
\rho = \sqrt{A + \frac{B}{a^6}},
\label{solution}
\end{equation}
where $B$ is an integration constant.

By choosing a positive value for $B$ we see that for small $a$
(i.e. $a^6 \ll B/A$)  the expression (\ref{solution}) is
approximated by
\begin{equation}
\rho \sim \frac{\sqrt{B}}{a^3} \label{dust}
\end{equation}
that corresponds to  a universe dominated by dust-like matter. For
large values of the cosmological radius $a$ it follows that
\begin{equation}
\rho \sim \sqrt{A}, \ p \sim - \sqrt{A},
\label{cosm}
\end{equation}
which, in turn, corresponds to an empty universe with a
cosmological constant $\sqrt{A}$ (i.e a de Sitter universe). In
the flat case it is possible also to find exact solutions as
follows:
\begin{eqnarray}
&& t = \frac{1}{6^4\!\!\sqrt{A}}\left(\ln
\frac{^4\!\!\sqrt{A+\frac{B}{a^6}}+^4\!\!\sqrt{A}}
{^4\!\!\sqrt{A+\frac{B}{a^6}}-^4\!\!\sqrt{A}} - 2\arctan
^4\!\!\sqrt{1+\frac{B}{Aa^6}}\right)\label{exact}
\end{eqnarray}

Note that $\sqrt A$ solves the equation \begin{equation} \rho +
p=\rho -\frac{A}\rho = 0. \end{equation} The circumstance that
this equation has a nonzero solution lies at the heart of the
possibility of interpreting the model as a ``quintessential''
model. If this model were realistic we could estimate the
constant $A$ by comparing our expressions for pressure and energy
with observational data. An indirect and naive way to do it is to
consider the nowadays accepted values for the contributions of
matter and cosmological constant to the energy density of the
universe. To use these data we decompose pressure and energy
density as follows:
\begin{eqnarray}
&& p = p_{\Lambda} + p_{M} = -\Lambda, \\
&& \rho = \rho_{\Lambda} + \rho_{M} = \Lambda + \rho_{M}.
\end{eqnarray}
An application of Eq. (\ref{Chapl}) gives
\begin{equation}
A = \Lambda (\Lambda + {\rho_{M}}). \label{param}
\end{equation}
If the cosmological constant contributes seventy percent of the
energy we get $\sqrt{A} \approx 1.2 \,\Lambda$. We now observe
that, in the context   of a Chaplygin cosmology, once an expanding
universe starts accelerating it cannot decelerate any more. Indeed
eqs. (\ref{Fried1}) and (\ref{Fried2}) imply that
\begin{equation}
\frac{\ddot{a}}{a} = -\frac12 (\rho + 3p). \label{Fried3}
\end{equation}
Condition $\ddot a>0$ is equivalent to
\begin{equation}
a^6 > \frac{B}{2A}, \label{cond}
\end{equation} which is obviously preserved by time evolution
in an expanding universe. It thus follows that the observed value
$\Lambda$ of the (effective) cosmological constant  will increase
up to $1.2 \, \Lambda$.

Considering now the subleading terms in  Eq. (\ref{solution}) at
large values of $a$ (i.e. $a^6 \gg B/A$), one obtains the
following expressions for the energy and pressure:
\begin{eqnarray}
&& \rho \approx \sqrt{A} + \sqrt{\frac{B} {4 A}}\ a^{-6},
\label{stiff}\\
&&p \approx -\sqrt{A} + \sqrt{\frac{B} {4 A}}\ a^{-6}.
\label{stiffpr}
\end{eqnarray}
Eqs. (\ref{stiff}) and (\ref{stiffpr}) describe the mixture of a
cosmological constant $\sqrt{A}$ with a type of matter known as
``stiff'' matter, and  described by the following equation of
state:
\begin{equation}
p = \rho.
\label{stiffeq}
\end{equation}
Note that a massless scalar field is a particular instance  of
stiff matter. Therefore, in a generic situation, a Chaplygin
cosmology can be looked at as interpolating between different
phases of the universe: from a dust dominated universe to a de
Sitter universe passing through an intermediate phase which is
the  mixture just mentioned above. The interesting point however
is that such an evolution is accounted by using only one fluid.

In recent series of paper \cite{parker} a similar type of
evolution   has been described, where the universe  passes from a
dust dominated epoch  to a de Sitter phase through an
intermediate phase  described as mixture of cosmological constant
and radiation. In \cite{parker}   the mechanism responsible for
this behaviour is  different from ours and is based on the quantum
corrections to the effective action of a massive scalar field.
However these corrections lead to a ``standard'' equation of state
in the form of a mixture. We can reproduce this type of evolution
by a slight modification to our ``exotic'' equation of state
(\ref{Chapl}), namely
\begin{equation}\label{new}
p = - \frac{C}{^3\!\!\!\sqrt{\rho}}
\end{equation}
The obvious generalization $p = -  C  {\rho}^{-\alpha}  $ (with
$\alpha \geq -1$) gives a similar evolution from dust to
cosmological constant with an intermediate epoch that can be seen
as a mixture of a cosmological constant with a fluid obeying the
state equation $p= \alpha \rho$.

For open or flat Chaplygin cosmologies ($K=-1,0$), the universe
always evolves from a decelerating to an accelerating epoch.  For
the closed Chaplygin cosmological models ($K=1$), the Friedmann
equations (\ref{Fried1}) and (\ref{Fried3}) say that it is
possible to have a static Einstein universe solution $ a_0 =
{(3A)}^{-\frac{1}{4}}$ provided the following condition holds:
\begin{equation}
B = \frac{2}{3\sqrt{3A}} \label{Einstein}
\end{equation}
When $ B > \frac{2}{3\sqrt{3A}} $ the cosmological radius $a(t)$
can take any value  while if $$ B < \frac{2}{3\sqrt{3A}} $$ there
are two possibilities: either
\begin{equation}
a< a_1 = \frac{1}{\sqrt{3A}} \left(\sqrt{3}\sin\frac{\varphi}{3} -
\cos\frac{\varphi}{3}\right) \label{amax}
\end{equation}
or
\begin{equation}
a> a_2 = \frac{2}{\sqrt{3A}}\cos\frac{\varphi}{3}, \label{amin}
\end{equation}
where $\varphi = \pi - \arccos  {3\sqrt{3A}B}/2 $. The region
 $a_1 < a < a_2$ is not accessible.

Results connected to those presented in this letter have been
obtained by Barrow \cite{barrow1,barrow2,barrow3,barrow4} who has
considered cosmologies with fluids admitting a bulk viscosity
proportional to a power of the density.

In the flat $K=0$ case the FRW equations for Chaplygin fits in
Barrow's scheme as a special case \cite{barrow2} and indeed a
transition from power law to exponential expansion is noticed
already in \cite{barrow1,Ellis}. However since the state equation
for the corresponding fluid is different from our
Eq.(\ref{Chapl}) this coincidence of the solutions is destroyed
by any small perturbation, for instance by a small spatial
curvature or by adding another matter source. We mention also that
the role of the Chaplygin-like behaviour in cosmology was also
noticed in Ref. \cite{Rosu,Rosu1}.

Following \cite{barrow3} we now try to find a homogeneous scalar
field $\phi(t)$ and a potential $V(\phi)$ to describe the
Chaplygin cosmology. Let us consider therefore the Lagrangian
\begin{equation}
L(\phi)= \frac{1}{2}\dot{\phi}^2 - V(\phi) \label{scalar}
\end{equation}
and set the energy density  of the field equal to that of the
Chaplygin gas:
\begin{equation}
\rho_\phi = \frac{1}{2}\dot{\phi}^2 + V(\phi) = \sqrt{A +
\frac{B}{a^6}}. \label{scalar1}
\end{equation}
The corresponding "pressure" coincides with the Lagrangian
density:
\begin{equation}
p_\phi = \frac{1}{2}\dot{\phi}^2 - V(\phi) = -\frac{A}{\sqrt{A +
\frac{B}{a^6}}}. \label{scalar2}
\end{equation}
It immediately follows that
\begin{equation}
\dot{\phi}^2 = \frac{B}{a^6 \sqrt{A + \frac{B}{a^6}}}
\label{scalar3}
\end{equation}
and
\begin{equation}
V(\phi) = \frac{2 a^6 \left(A + \frac{B}{a^6}\right) - B}{2
a^6\sqrt{A + \frac{B}{a^6}}}. \label{scalar4}
\end{equation}
Now let us restrict ourselves to the flat case $K=0$. In this
case Eq. (\ref{scalar3}) also implies that
\begin{equation}
\phi' = \frac{\sqrt{B}}{a(A a^6 + B)^{1/2}}, \label{scalar5}
\end{equation}
where prime means differentiation w.r.t. $a$. This  equation can
be integrated and it follows that
\begin{equation}
a^6 = \frac{4B\exp(6\phi)}{A (1 - \exp(6\phi))^2}. \label{scalar6}
\end{equation}
Finally, by substituting the latter expression for the
cosmological radius in Eq. (\ref{scalar4}) one obtains the
following potential, which has a surprisingly simple form:
\begin{equation}
V(\phi) = \frac{1}{2}\sqrt{A}\left( \cosh 3\phi + \frac{1}{\cosh
3\phi}\right). \label{potential}
\end{equation}
Note that the potential does not depend on the integration
constant $B$ and therefore it reflects only the state equation
(\ref{Chapl}) as it should.

In conclusion Chaplygin cosmology provides an interesting
possibility to account for current observations about the
expansion of the universe. It predicts also that the cosmological
constant will increase (or that it was less in the past) and this
could in principle be observed. Of course to take this model
seriously one should have a good fundamental reason to believe in
Eq. (\ref{Chapl}).

{\bf Acknowledgments}: We are grateful to Prof. J.D. Barrow who
has kindly attracted our attention to his papers
\cite{barrow1,barrow2,barrow3,barrow4}. The work of A.K. was
partially supported by RFBR via grants No 99-02-18409 and No
99-02-16224 and by the Cariplo Science Foundation.

\end{document}